# Assessment of trends in an integrated climate metric - Analysis of 200 mbar zonal wind for the period 1958–2021

Nenghan Wan[1], Xiaomao Lin[1*], Roger A. Pielke Sr. [2*]

[1] Department of Agronomy, Kansas Climate Center, Kansas State University, Manhattan, KS, USA

[2] Cooperative Institute for Research in Environmental Sciences, University of Colorado, Boulder, CO, USA

Corresponding author: Roger A. Pielke Sr. (pielkesr@gmail.com), Xiaomao Lin (xlin@ksu.edu)

**Key Points:**

- The ERA5 and JRA-55 reanalysis provide closely correlated interannual and multidecadal trends in the 200 hPa zonal wind speeds. Over the period of 1958-2021, there is essentially no change in either the JRA-55 or ERA5 200 hPa zonal wind speeds.

- The polar jet core has moved slightly poleward in the Northern Hemisphere, but the movement is not statistically significant.

- No poleward shift in the polar jet core has occurred in the Southern Hemisphere.

**Abstract**

Using the ERA5 and JRA-55 reanalysis datasets on latitude bands of nearly equal area, we analyze the trend of 200 mbar zonal wind anomalies for the period 1958-2021 as well as annual-mean latitude of the polar jet stream. We find basically no change of 200 mbar zonal wind in either the JRA-55 or ERA5 dataset. In addition, the polar jet core moves poleward in the Northern Hemisphere, but the movement is not statistically significant, while no poleward shift at all has occurred in the Southern Hemisphere.

**Plain Language Summary**

The gradient wind between Earth's surface to the height of the 200 mbar is an effective integrator to relate the thickness between pressure surfaces to depth integrated temperatures. An increase in wind speeds, for example, occurs if this layer warmed at lower latitudes and/or cooled at higher latitudes, and/or the horizontal difference in temperature occurred over a shorter distance. We analyze the trend of gradient wind at east-west direction from 1958 to 2021 as well as latitudinal region with the strongest winds using two reanalysis datasets and found no change of gradient wind. In addition, the location of the strongest winds moves poleward in the Northern Hemisphere, but the movement is not statistically significant, while no poleward shift at all has occurred in the Southern Hemisphere.

**1 Introduction**

The assessment of the global average surface temperature as the main metric to monitor changes in climate has become a major international policy measure. The Paris Agreement of holding the global average surface temperature change to well below 2 °C and limiting it to 1.5 °C has become an international policy goal (Rogelj et al., 2016). However, the surface temperature data suffers from large spatial variations in coverage, biases (Karl et al., 2015) from site exposure and representativeness etc., which must be accounted for (Pielke et al., 2007). Using spatially heterogeneous and relatively sparse (in many regions) data is also applying a two-dimensional dataset to characterize the three-dimensional climate system. Rather, it is much more robust to analyze the three-dimensional atmosphere. Indeed, it is optimal to let the atmosphere itself do the averaging.

There are particular weather observations in which the atmosphere itself performs the integration best suited to evaluate trends. As is well known, integration of data reduces the effect of random errors, although systematic errors, if any, would still persist. For mid- and high latitudes, the gradient wind relationship can be used to quantitatively relate the thickness between pressure surfaces to depth-integrated temperatures (Pielke, 2013). The gradient wind is the same as the geostrophic wind when horizontal curvature of the flow is relatively small. Even in low latitudes, this relationship is reasonably accurate as winds are still closely related to the gradient winds at larger spatial scales.

Previous studies have successfully used zonal wind as a zonal mean temperature proxy applying the thermal wind equation with reanalysis, radiosonde data, and pilot balloons to characterize warming (Allen & Sherwood, 2008; Pralungo & Haimberger, 2015; Lee et al., 2019). In Pielke et al. (2001), the 200 mbar zonal winds ($u$ component) for the period 1958–1997 were analyzed using NCEP reanalysis data. From the thermal wind relation, the winds at 200 mbar (which are very close to the gradient wind speed) are an effective integrator of the tropospheric meridional temperature gradient averaged from the surface up to the altitude of the 200 mbar pressure surface. The monitoring of trends in the 200 mbar winds, therefore, provides an approach to examine long-term changes and variability in the atmospheric circulation of the troposphere.

The strong westerly winds generated as a consequence of the equator-to-pole temperature gradient near the tropopause form the polar jet stream. Changes in the location, intensity, or altitude of the polar jet stream, can influence mid-latitude weather systems as well as air transport (Hannachi et al., 2012; Williams, 2016). Francis and Vavrus (2012, 2015) concluded that a wavier polar jet stream in combination with slowing of upper-level zonal wind due to rapid Arctic warming, increases the likelihood of persistent weather patterns and extreme events. Williams (2016) reported a strengthening of the prevailing jet stream winds causes eastbound flights over the north Atlantic to significantly shorten and westbound flights to significantly lengthen in all seasons. Eastbound and westbound crossings in winter become approximately twice as likely to take under 5 hours 20 minutes and over 7 hours 00 minutes, respectively, but this study did not take a global perspective.

In this paper, the 200 mbar zonal wind was selected as an effective way to assess variability and trends in atmospheric circulation based on the thermal wind relationship (discussed in detail in the next section). We analyze the trend of the 200 mbar zonal wind ($u$ component) anomaly for the period 1958-2021 as well as the annual mean latitude of the polar jet stream using ERA5 and JRA-55 reanalysis datasets at the global level. Data and methodology are explained in section 3. Zonal wind trends are presented in section 4. Finally, conclusions are presented in section 5.

## 2 Thermal Wind Relationship

The dynamical relationship known as the thermal wind equation is used to calculate differences in the geostrophic wind between the surface (1000 mbar level in this study) and the 200 mbar level. The component of the thermal wind equation in the east–west wind speed ($u$) in terms of the layer-averaged temperature gradient is expressed as Eq.1 (Bluestein, 1992; R. A. Pielke et al., 2001).

$$u = \frac{R_d}{f} \ln\left(\frac{1000 \; mbar}{200 \; mbar}\right) \left(k \times \nabla_p \bar{T}\right) \cdot i \qquad (1)$$

Where $f$ is the Coriolis parameter, $R_d$ is the gas constant for dry air, $\nabla_p \bar{T}$ is the horizontal temperature gradient averaged from 1000 to 200 mbar. The gradient operator $\nabla_p$ denotes a gradient on an isobaric level. $k$ and $i$ are the vertical unit vector and horizontal (east-west) unit vector. The westerly winds at 200 mbar, $u$200 mbar, are thus directly related to the magnitude of the horizontal, north-south gradient in temperature averaged from the Earth's surface to the height of the 200 mbar winds where the surface geostrophic wind (the surface is defined to be at 1000 mbar) is assumed to be negligible compared to the 200 mbar winds. The latitudinal region with the strongest 200 mbar winds is called the polar jet stream.

An increase in wind speeds, for example, occurs if this layer warmed at lower latitudes and/or cooled at higher latitudes, and/or the horizontal difference in temperature occurred over a shorter distance. An increase in the layer-mean horizontal temperature gradient of 1 °C per 1000 km at a latitude of 43 degrees, for example, would produce a 200 mbar gradient wind speed increase of

4.6 m s$^{-1}$; an effect which should be easily detectable in observational data. An advantage of using (1) is that the atmosphere itself performs the vertical integration of the layer-averaged horizontal temperature gradient. The accuracy of (1) has been repeatedly confirmed through independent calculations of the term delta T and comparison with observed winds (Pielke, 1988).

## 3 Data source and method

The fifth generation ECMWF atmospheric reanalysis of the global climate (ERA5) dataset is the replacement for ERA-Interim reanalysis. Benefiting from developments in model physics, core dynamics, and data assimilation, this dataset has the advantage of much higher temporal and spatial resolution (Hersbach et al., 2020). Currently data is available from 1950 and has two versions for 1950-1978 (preliminary back extension) and 1979 onwards (final release plus timely updates). We collected monthly-averaged data from 1958 to 2021 on a 1440×721 grid with 0.25° latitude and 0.25° longitude grid increments.

The Japanese 55-year Reanalysis (JRA-55) is the second Japanese global atmospheric reanalysis commissioned by the Japan Meteorological Agency (JMA) which covers the period start from 1958 (Kobayashi et al., 2015). It is the upgraded version of JRA-25 and has undergone a variety of improvements including higher spatial resolution and a new radiation scheme. Monthly-averaged data were chosen on a grid of 288×145 points with a horizontal spacing of 1.25°×1.25° from 1958 to 2021.

We analyzed the 200 mbar annual westerly wind anomaly defined as a departure from the base-period year average (from 1958 to 1997) for the period 1958-1997 and 1998-2021 from ERA5 and JRA-55 monthly-averaged datasets. All datasets were used on a standard latitude-longitude grid. Linear trends were calculated from ordinary least-squares regression and statistical significance was tested at the 95% confidence level ($p<0.05$) according to a two-tailed t-test.

The mean of the westerly wind anomaly was calculated with a cosine (latitude) weighting factor to account for the convergence of grid points at near-equal area latitude bands that approximately account for 4% of Earth areas for each 0° to 5°, 5° to 10°, 10° to 15°, 15° to 20°, 20° to 26°, 26° to 32°, 32° to 38°, 38° to 45°, 45° to 52°, 52° to 61° and 61° to 73° bands in the Southern and

Northern Hemisphere. The 73°-90° bands in the Southern and Northern Hemisphere are defined separately and account for approximately 2% of the Earth's area. We also calculated trends in the annual-mean 200 mbar wind speed between 20°-60° latitude for the Northern and Southern Hemisphere in a similar approach to that used in Bracegirdle et al. (2018). At each latitude, the all-longitude average was calculated for each month, and then all the locations of maximum wind speed in each month were averaged for each year. The averaged latitude was then selected as the polar jet core location.

## 4 Results

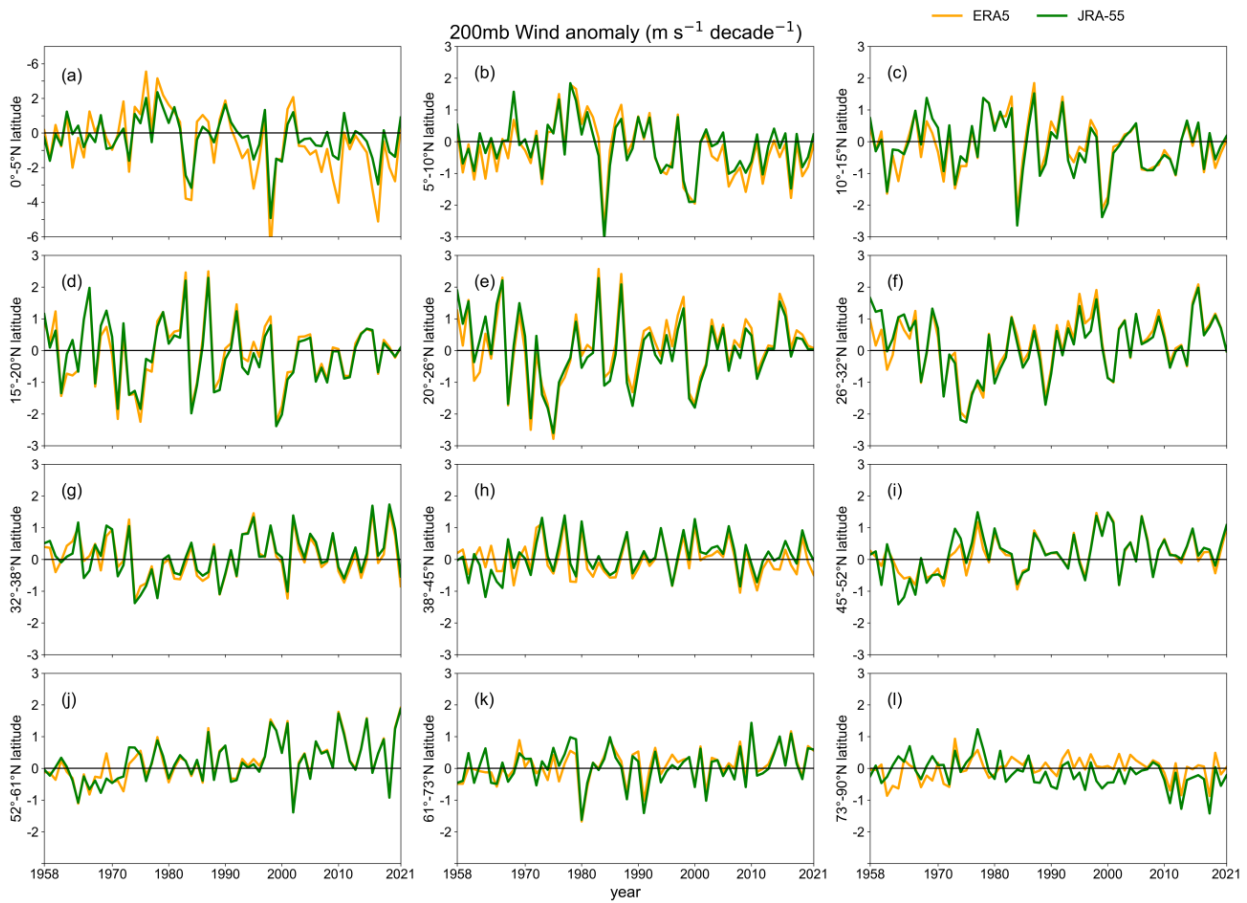

**Figure 1.** The 200 mbar westerly wind anomaly in meters per second per decade from the ERA5 (orange line) and JRA-55 (green line) for 1958–2021 in the Northern Hemisphere. (a) 0°–5°N, (b) 5°–10°N, (c) 10°–15°N, (d) 15°–20°N, (e) 20°–26°N, (f) 26°–32°N, (g) 32°–38°N, (h) 38°–45°N, (i) 45°–52°N, (j) 52°–61°N, (k) 61°–73°N, and (l) 73°–90°N.

Averaged over the region of near-equal area latitude bands in the Northern Hemisphere, the timeseries of the 200 mbar zonal wind anomalies since 1958 are very similar except for a slight difference in magnitude in the 0°-5°N band. Both ERA5 and JRA-55 reanalysis datasets are in good agreement with respect to the inner-annual variability. Since 1958, the latitude bands 0°-5°N, 5°-10°N, 10°-15°N, and 15°-20°N show a decrease in westerly wind speed in both the ERA5 and JRA-55 reanalysis datasets, but only the trends in 0°-5°N and 5°-10°N latitude bands are statistically significant at the 95% confidence level for the ERA5 dataset (Table 1).

The trend shows an increase between 26°-32°N and 32°-38°N as well as 45°-52°N and 52°-61°N, however only the 45°-52°N and 52°-61°N latitude bands had a significant increase in the westerly anomaly trend at a rate of approximately 0.1 m s$^{-1}$ decade$^{-1}$ in both the ERA5 (p<0.05) and the JRA-55 (p<0.05) reanalysis. There are some discrepancies in latitude bands 20°-26°N, 38°-45°N, and 73°-90°N, with ERA5 showing a decrease of zonal wind in the 38°-45°N and an increase in 20°-26°N and 73°-90°N, while JRA-55 is opposite (Figs.1(e), (h), (l)).

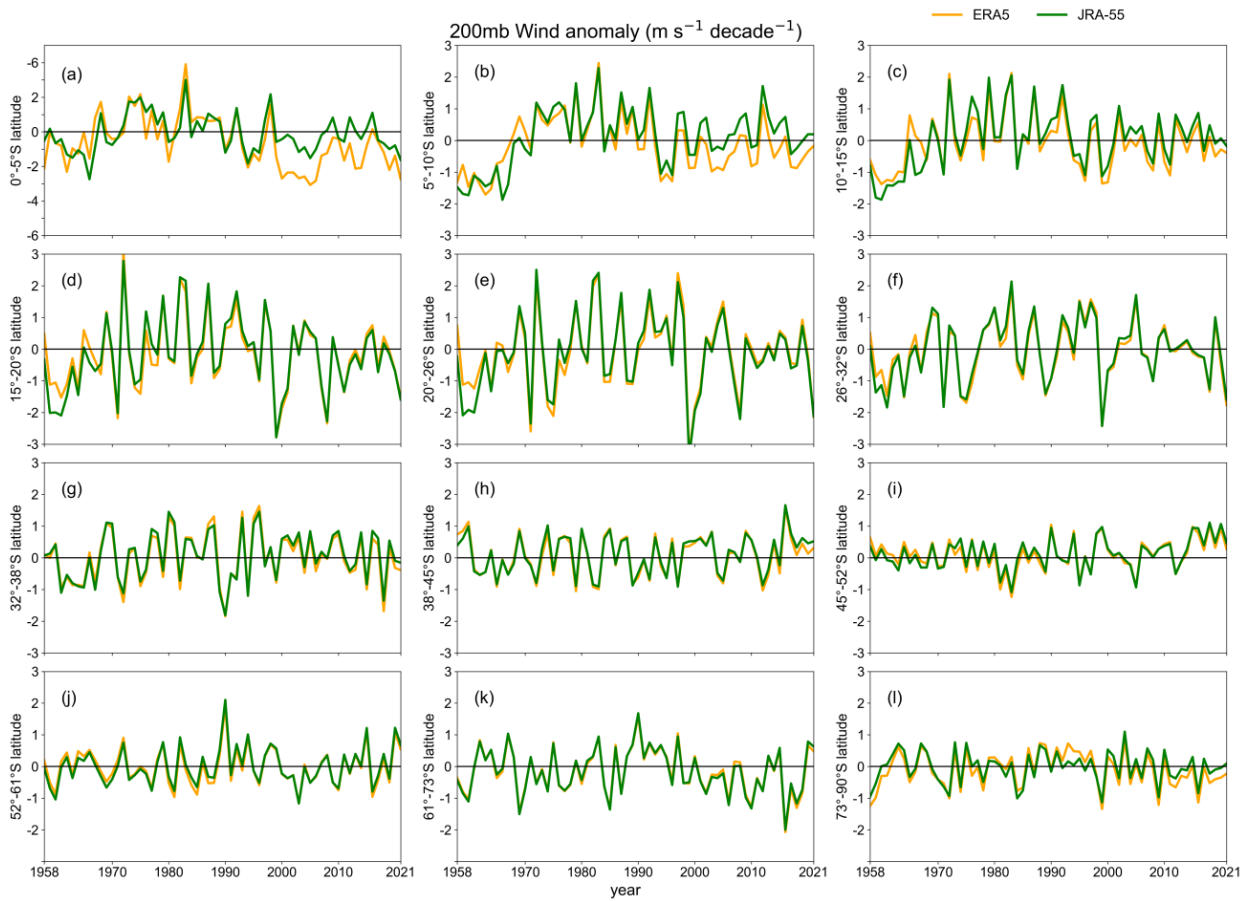

**Figure 2.** The 200 mbar westerly wind anomaly in meters per second per decade from the ERA5 (orange line) and JRA-55 Reanalysis (green line) for 1958–2021 in the Southern Hemisphere. (a) 0°–5°S, (b) 5°–10°S, (c) 10°–15°S, (d) 15°–20°S, (e) 20°–26°S, (f) 26°–32°S, (g) 32°–38°S, (h) 38°–45°S, (i) 45°–52°S, (j) 52°–61°S, (k) 61°–73°S, and (l) 73°–90°S.

Similarly, the time series of the 200 mbar zonal wind anomalies is provided in Figure 2 for the Southern Hemisphere since 1958, averaged over the region of near-equal area latitude bands. The agreement of the two reanalysis datasets is obviously better than that in the Northern Hemisphere as illustrated in their trends shown in Table 1. The wind speeds decrease in the 0°-5°S, 61°-73°S, and 73°-90°S latitude bands, but only the trend in 0°-5°S of the ERA5 dataset shows a statistically significant decrease. The trend increases in 10°-15°S as well as 26°S to 61°S latitudes while only the values in 10°-15°S and 45°-52°S for JRA-55 dataset are significant at the 95% confidence level.

**Table 1.** *200 mbar observed westerly wind linear trend in meters per second per decade across different latitude bands from the ERA5 and JRA-55 reanalysis datasets, * represents $p < 0.05$.*

| | 1958-1997 | | 1998-2021 | | 1958-2021 | |
|---|---|---|---|---|---|---|
| Latitude bands | ERA5 trend | JRA-55 trend | ERA5 trend | JRA-55 trend | ERA5 trend | JRA-55 trend |
| 0°-5°N | -0.12 | 0.06 | -0.10 | 0.33 | -0.37* | -0.13 |
| 5°-10°N | 0.11 | -0.03 | 0.23 | 0.32 | -0.12* | -0.10 |
| 10°-15°N | 0.15 | -0.03 | 0.20 | 0.36 | -0.04 | -0.07 |
| 15°-20°N | 0.08 | -0.08 | 0.36 | 0.46 | -0.01 | -0.06 |
| 20°-26°N | 0.07 | -0.20 | 0.36 | 0.38 | 0.08 | -0.02 |
| 26°-32°N | 0.08 | -0.14 | 0.20 | 0.22 | 0.14* | 0.07 |
| 32°-38°N | -0.05 | -0.03 | 0.02 | 0.07 | 0.03 | 0.07 |
| 38°-45°N | -0.04 | 0.08 | -0.26 | -0.17 | -0.03 | 0.06 |
| 45°-52°N | 0.10 | 0.16 | -0.28 | -0.19 | 0.10* | 0.13* |
| 52°-61°N | 0.15* | 0.13 | 0.06 | 0.07 | 0.18* | 0.16* |
| 61°-73°N | 0.09 | -0.01 | 0.22 | 0.24 | 0.09* | 0.05 |
| 73°-90°N | 0.13* | -0.08 | -0.14 | -0.09 | 0.02 | -0.10 |
| 0°-5°S | 0.13 | 0.18 | 0.13 | -0.22 | -0.30* | -0.03 |
| 5°-10°S | 0.27* | 0.48* | 0.04 | 0.04 | -0.01 | 0.17* |
| 10°-15°S | 0.24 | 0.45* | 0.08 | 0.08 | 0.01 | 0.13* |
| 15°-20°S | 0.30 | 0.51* | 0.18 | 0.09 | -0.01 | 0.03 |
| 20°-26°S | 0.34* | 0.49* | 0.12 | 0.17 | 0.03 | 0.04 |
| 26°-32°S | 0.28* | 0.38* | -0.14 | -0.10 | 0.03 | 0.06 |
| 32°-38°S | 0.16 | 0.13 | -0.24 | -0.17 | 0.03 | 0.06 |
| 38°-45°S | -0.08 | -0.06 | -0.01 | 0.08 | 0.02 | 0.06 |
| 45°-52°S | -0.10 | -0.03 | 0.16 | 0.22 | 0.05 | 0.08* |
| 52°-61°S | 0.03 | 0.14 | 0.14 | 0.21 | 0.01 | 0.05 |
| 61°-73°S | 0.19 | 0.19 | 0.01 | 0.10 | -0.04 | -0.04 |
| 73°-90°S | 0.20* | 0.04 | -0.09 | -0.02 | -0.02 | 0.00 |

Table 1 shows the trend of different latitude bands from 1958-1997, 1998-2021, and overall. From 1958 to 1997, ERA5 and JRA-55 show a good agreement with the increase in zonal wind for 45°-52°N and 52°-61°N, while only the trend in the band 52°-61°N is statistically significant (p<0.05). In the Southern Hemisphere, wind significantly increases at around a rate of 0.4 m s$^{-1}$ decade$^{-1}$ in the JRA-55 dataset (p<0.05) over 5°-32°S and decreases for 38°-45°S and 45°-52°S but does not show a statistical significance. From 1998 to 2021, all trends in the Southern and Northern Hemisphere of both datasets are in good agreement except the 0°-5° latitude bands. Wind increases over the 5°- 26° latitude band as well as the 52°-73° latitude band in both Hemispheres, but decreases in the 26°-32°S, 32°-38°S, 38°-45°N, and 45°-52°N bands. Comparing the two periods, the trend shifts from a significant increase to a decrease in 26°-32°S and 32°-38°S while the wind increases over two periods from 5°-26°S latitude.

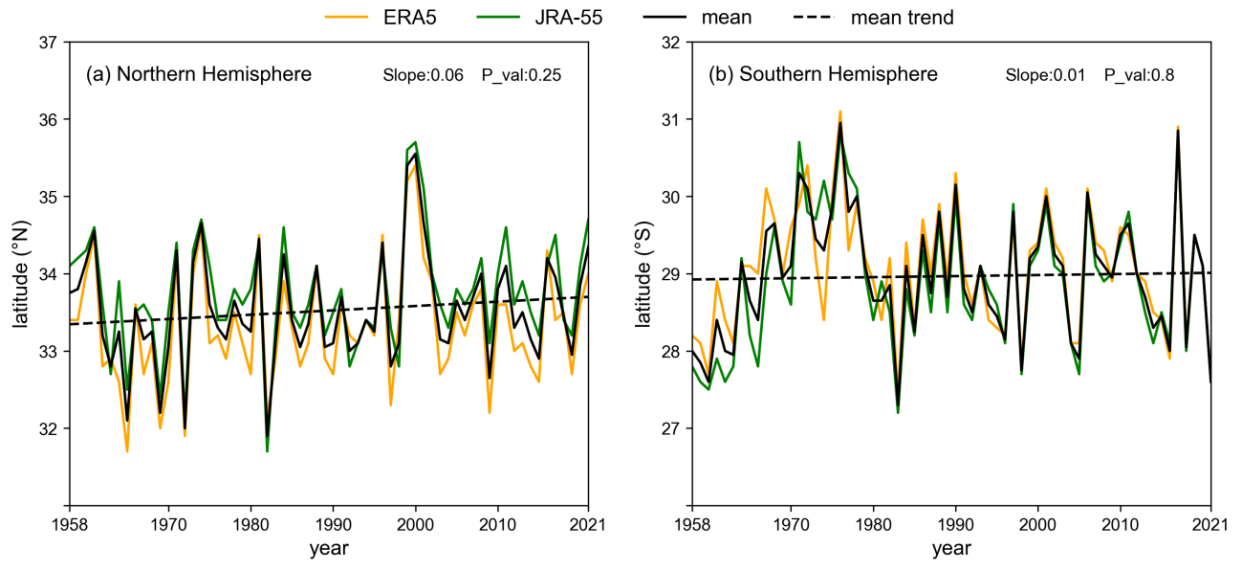

**Figure 3.** Annual mean latitude of the maximum zonal wind speed from the ERA5 (orange line) and JRA-55 Reanalysis (green line) and the linear trend in the mean (black line) for 1958–2021

in the (a) Northern and (b) Southern Hemisphere.   The positive trend indicates a poleward shift in the jet core while the slope is shown in degrees decade$^{-1}$.

Using the maximum 200 mbar zonal wind speed, a jet core location can be obtained. We calculated the annual mean latitude of maximum zonal wind speed between 20°-60° over the Southern and Northern Hemispheres for the period 1958-2021 from both the ERA5 and JRA-55 datasets. In Figure 3 (a), we found the jet core moved poleward in the Northern Hemisphere during this time period at an average rate of 0.06 degrees decade$^{-1}$ but the trend was not statistically significant for either dataset. This trend agrees with previous studies although the methods used are different and the trend is not as strong as found in Molnos et al. (2017) and Rikus (2018). The trend in the Southern Hemisphere also shows a poleward shift but with no statistically significant trend (Fig. 3 (b)) although it has the same sign as in Archer and Caldeira (2008) and Pena-Ortiz et al. (2013) but a smaller value.

## 5 Conclusions

The ERA5 and JRA-55 reanalysis provide the most accurate tool available to assess long-term changes in the tropospheric wind and temperature fields. Adding to this tool, we can utilize the fundamental concept in atmospheric physics that relates large-scale wind speed in the troposphere to the spatial gradient of the vertically-integrated temperatures below that level. We focus on the east-west component of the wind at 200 hPa in order to assess the north-south gradient in layer-averaged temperatures below that level. For example, a larger spatial gradient will produce stronger winds at 200 hPa resulting in a much stronger winter polar jet than observed in the summer.

Using latitude bands of near-equal area, only a few trends are statistically significant. The largest trend is 0.51 m s$^{-1}$ per decade in the 15°S-20°S bin in JRA-55 from 1958-1997, although not in the equivalent ERA5 reanalysis. Remarkably, over the entire period of 1958-2021, there is essentially no change in either the JRA-55 or ERA5.  The polar jet core does show migration to the north in the Northern Hemisphere at about 0.5 degrees of latitude, but the change is not statistically significant. The Southern Hemisphere has essentially no trend. These findings document only small changes to the 200 hPa wind speeds from 1958-2021.

One explanation for the lack of statistically significant trends is based on previous studies (e.g., (Chase et al., 2002; Tsukernik et al., 2004; Herman et al., 2008). There is a constraint on warming in the tropical mid-troposphere and cooling in the winter in the higher latitude mid-troposphere due to deep cumulus convection which limits the warmest 500 hPa temperatures to -3 °C and the coldest 500 hPa temperatures to -42 °C (Chase et al., 2015). These coldest temperatures are achieved before the winter solstice occurs. The 500 hPa temperatures are accepted as the optimal individual tropospheric level to characterize tropospheric dynamics including tropospheric layer-averaged temperatures (Pielke Sr, 2013). This constraint on the tropospheric layer-averaged temperatures, therefore, constrains wind speeds at 200 hPa.

This has implications for claims that the polar jet has become weaker (and wavier) due to a warming Arctic troposphere (Francis & Vavrus, 2015). Our analysis contradicts that view. The polar jet is as strong in 2021 as it was in 1958. Since the polar jet stream results in extratropical cyclone development through baroclinic development (Holton, 1973; Pielke, 2013), a lack of a significant trend in the polar jet and in the 200 hPa winds, indicates that these cyclones should be as frequent and intense in 2021 as they were in 1958.

**Acknowledgments**
This study was supported in part by the U.S. Department of Agriculture, National Institute of Food and Agriculture (grant no. 2016-68007- 25066). The contribution number of this manuscript is 22-276-J. We thank Dallas Staley for her outstanding contribution in editing and finalizing the paper. Her work continues to be at the highest professional level.

**Data Availability Statement**
ERA5 data are available to download for registered users at Copernicus Data Store: monthly data on pressure level from 1950 -1979 are available at https://cds.climate.copernicus.eu/cdsapp#!/dataset/reanalysis-era5-pressure-levels-monthly-means-preliminary-back-extension?tab=overview. and monthly data on pressure level from 19790 onwards are available at https://cds.climate.copernicus.eu/cdsapp#!/dataset/reanalysis-

era5-pressure-levels-monthly-means?tab=overview.  NCAR and Japan Meteorological Agency data for JRA-55 is available at https://doi.org/10.5065/D60G3H5B. Accessed: 24 Mar. 2022.